\documentclass{owrart}

\usepackage{color}

\begin{document}

%% --------------------------------------------------------------------------
%% Please use the environment "talk" for each abstract.
%% It has three obligatory and one optional argument. The syntax is:
%% -----------------------
%% \begin{talk}[coauthors]{Name of the speaker}{Title of the talk}{Author Sorting Index}
%%      .....
%% \end{talk}
%% -----------------------
%% The names of coauthors will appear in form of "(joint work with ...)"
%%
%% The Author Sorting Index should be given as the last and first name of the speaker,
%% separated by a comma. If for example the name of the speaker is "John Smith", then
%% the correct Author Sorting Index is "Smith, John".
%% Any special characters (like accents, German umlaute, etc.) should be replaced by
%% their "non-special" version, eg replace \"a by a, \'a by a, etc.
%%
%% Please use the standard thebibliography environment to include
%% your references, and try to use labels for the bibitems, which
%% are uniquely assigned to you in order to avoid conflicts with other authors.
%% You can achieve unique labels by using our on initials before every label.
%% -------------------------------------------------------------------------------

\begin{talk}[]{Daniela Cadamuro}
{Wedge-local fields in interacting quantum field theories with bound states}
{Cadamuro, Daniela}

\noindent
The construction of interacting quantum field theories is a hard task due to the complicated structure of local observables in the presence of interaction. In the class of \emph{quantum integrable models} in $1+1$-dimensional Minkowski spacetime the construction becomes more tractable due to the particular type of interaction. Integrable models describe systems of relativistic particles   subject to elastic scattering, where the momenta of the particles and the particle number are conserved, yielding infinitely many conserved quantities. As a consequence, the $S$-matrix is of ``factorizing'' type in the sense that the scattering of $n$ particles is the product of two particle scattering processes. 

These models have been treated with various methods. Using the perturbative approach, one computes the $S$-matrix elements, and therefore the Green's functions, of the theory. However, Lagrangians are usually complicated enough that the construction remains at perturbative level. An exception is the sine-Gordon model where Fr\"ohlich and Seiler \cite{FS76} compute the Euclidean Green's functions and show that the $S$-matrix is non-trivial. However, it has not been proven that the $S$-matrix is factorizing. Alternatively, the \emph{Form Factor Programme} \cite{BK04, BFK06} bases the construction on an inverse scattering problem. One postulates the form of the S-matrix and, via a list of axioms deduced from physical requirements and consistent with the properties of the S-matrix, constructs the $n$-point functions of the theory by expanding them in a series of \emph{form factors} (i.e., certain matrix elements of the interacting fields). These yield infinite series expansions whose convergence is hard to control.  

A more recent idea due to Schroer \cite{S99} avoids these infinite series by considering, instead of strictly local operators, observables localized in unbounded wedge-shaped regions. This weaker localization property allows to construct observables with a simpler expression in momentum space. Strictly local observables can then be recovered by taking intersection of the algebras generated by observables in \emph{right} and \emph{left} wedges. Using an abstract argument based on a certain phase space property called \emph{modular nuclearity}, one would show that this intersection is non-trivial. Finally, one would solve the inverse scattering problem by computing the $S$-matrix of the input by methods of Haag-Ruelle scattering theory. This \emph{operator-algebraic approach} has proved to be successful for the construction of a large class of integrable models. These are models of bosons with an $S$-matrix analytic in the \emph{physical strip} (a certain region of the momentum complex plane), including the Ising and sinh-Gordon models with one particle species \cite{L08}, Federbush-type models \cite{T14} with several particle species; there are also partial results in the $O(N)$ non-linear sigma models \cite{A14,LS14}.

However, models where $S$-matrix components have singularites in the physical strip have not been treated in this framework before. In \cite{CT15, CT16}, together with Y.~Tanimoto, I obtained first results  in this direction. The models we consider have $S$-matrices whose components have a certain pole structure in the physical strip. Examples are the Bullough-Dodd model, the $Z(N)$-Ising model, the affine-Toda field theories, the sine-Gordon and Thirring models. These models are of interest since a pole in the physical strip 
is believed to correspond to a bound state. The idea is that two particles of type $\alpha, \beta$ can scatter with the exchange of a unitary factor, the $S$-matrix component $S^{\alpha \beta}_{\beta \alpha}(\theta_1 - \theta_2)$, but the type of particles stays the same in scattering. In this case, the $S$-matrix is said to be ``diagonal''. 
We denote with $\theta_1, \theta_2$ the \emph{rapidities} of the particles $\alpha, \beta$, which parametrize their momenta. The two particles can also fuse into a third particle of type $\gamma$ in the following sense:
\begin{equation}\label{conservation}
p_{m_\alpha}(\theta +i\theta_{(\alpha \beta)}) + p_{m_\beta}(\theta -i\theta_{(\beta \alpha)}) = p_{m_\gamma}(\theta),
\end{equation}
i.e.,  the momenta of two “virtual” particles add to the momentum of a third “real” particle (the bound particle) which lies on the mass shell.  The numbers $\theta_{(\alpha \beta)}, \theta_{(\beta \alpha)}$ are solutions of this equation, once the masses of the particles are fixed.
The bound state formed would correspond to a pair of simple poles of the  component $S^{\alpha \beta}_{\beta \alpha}(\zeta)$ in the physical strip $0<\operatorname{Im} \zeta <\pi$. These are the \emph{s-channel} pole, which is located at $\zeta= i\theta_{\alpha \beta} = i\theta_{(\alpha \beta)} +i \theta_{(\beta \alpha)}$ and the \emph{t-channel} pole, which arises due symmetry properties of the S-matrix.
The possible fusion processes, that we denote by the symbol $(\alpha \beta) \rightarrow \gamma$, are characteristic of each model and they form their fusion tables. Here, we will focus on the $Z(N)$-Ising model, where the particles are of type $1, \ldots, N-1$, the possible fusions are of the form $(\alpha \beta) \rightarrow \alpha + \beta \operatorname{mod} N$, and the anti-particle $\bar \alpha$ of a particle of type $\alpha$ is $N-\alpha$.   

For each pair $\alpha, \beta$, the component $S^{\alpha \beta}_{\beta \alpha}(\zeta)$ of the $S$-matrix is a meromorphic function on $\mathbb{C}$, fulfilling a number of axioms \cite[Sec.~2.1]{CT16}, including unitarity, crossing symmetry and the Bootstrap equation. 
As an example,  the crossing symmetry relation reads $S^{\alpha \beta}_{\beta \alpha}(i\pi - \zeta) = S^{\bar \beta \alpha}_{\alpha \bar \beta}(\zeta)$, for $\zeta \in \mathbb{C}$.
In the $Z(N)$-Ising model, the component $S^{11}_{11}(\zeta)$, fulfilling the above properties, is given by
\begin{equation}
S^{11}_{11}(\zeta) = \frac{\sinh \frac{1}{2}\big( \zeta + \frac{2\pi i}{N} \big)}{ \sinh \frac{1}{2}\big( \zeta - \frac{2\pi i}{N} \big)},
\end{equation}
and one can construct all the other $S$-matrix components by using the Bootstrap equation. Specfically, the components $S^{\alpha \beta}_{\beta \alpha}$ with only indices of type $1$ and $N-1$ have at most two simple poles in the physical strip and no other poles. These components play a crucial role in the proof of \emph{weak wedge-commutativity}, as we will explain below.

For an $S$-matrix $S^{\alpha \beta}_{\beta \alpha}$ analytic in the physical strip, Lechner constructed quantum fields $\phi, \phi'$ with the property that they commute when smeared with test functions supported in the left and right wedge, respectively.  This computation relies on a shift of an integral contour where the integrand contains $S^{\alpha \beta}_{\beta \alpha}$. In the case where the $S$-matrix has poles in the physical strip, shifting the integral contour yields the residues of $S^{\alpha \beta}_{\beta \alpha}$, and $\phi, \phi'$ are no longer wedge-local. To overcome this problem, we introduce a new field $\tilde \phi = \phi + \chi$ by adding the so called \emph{bound state operator} $\chi$ to the field $\phi$, so that the commutator of $\chi$ with its reflected operator (by the action of the CPT operator) $\chi'$ cancels the above residues. $\chi$ acts on a one-particle vector $\xi^\beta$ as follows, if $(\alpha \beta)$ fuse into some $\gamma$,
\begin{equation}
\big( \chi_{1,\alpha}(f) \xi \big)^\gamma (\theta) := -i\eta^\gamma_{\alpha \beta} f^+_{\alpha}(\theta +i\theta^\gamma_{(\alpha \beta)}) \xi^\beta(\theta -i\theta_{(\beta \alpha)}),
\end{equation}
where the matrix elements $\eta^\gamma_{\alpha \beta}$ are related to the residues of the $S$-matrix. One can show that this operator is densely defined and symmetric on a suitable domain of vectors, but it is not self-adjoint on a naive domain. To prove (weak) wedge-commutativity, one shows that the commutator of the reflected field $\tilde \phi '$ with $\tilde \phi$ vanishes for test functions supported in right and left wedges, respectively, in the weak sense, i.e. in matrix elements between suitable vectors. In this computation in the $Z(N)$-Ising model, the components of the test functions, and therefore the $S$-matrix components, are restricted to particles of type $1$ and $N-1$, limiting the number of poles in the physical strip. Strong commutativity has not been proven yet. 
For this, one would need to show the existence of self-adjoint extensions of $\tilde \phi(f)$ and $\tilde \phi'(g)$, and  select the ones that strongly commute. This is a non-trivial task, but some progress has been made in the Bullough-Dodd model. $\tilde \phi(f)$ is also a polarization-free generator but not temperate. In the case of many particle species, we can prove the Reeh-Schlieder property only for models with two species of particles. 

In the operator-algebraic approach, we are interested in the following question. Let us suppose that strong commutativity hold for a certain extension $\tilde \phi^-$. We consider the von Neumann algebras
\begin{equation}
\mathcal{A}(W_L + x) = \{  e^{i\tilde \phi(f)^-} : \operatorname{supp} f \subset W_L + x  \}^{''},
\end{equation}
and similarly for the right wedge. Observables localized in bounded regions are obtained as intersections of von Neumann algebras
\begin{equation}
\mathcal{A}(\mathcal{O}) := \mathcal{A}(W_L +x) \cap \mathcal{A}(W_R -y), \quad \mathcal{O} = W_L + x \cap W_R -y.
\end{equation}
The problem would then be to show that this intersection is non-trivial, i.e., technically, that the vacuum of the theory is cyclic and separating for the local algebra.  Finally, \emph{Haag-Ruelle scattering theory} would be applied to compute the $S$-matrix of the input and solve the inverse scattering problem. This would yield a complete construction of the theory, and is part of our future work. 

Concluding, the construction of integrable models with bound states is a new promising direction in constructive  quantum field theory. The Bullough-Dodd model, the $Z(N)$-Ising model and affine-Toda field theories are among those models which we hope to fully construct using operator-algebraic techniques. An interesting problem would be to extend our construction to the sine-Gordon  and Thirring models, comparing our construction to the Euclidean methods in \cite{FS76}. It would also be interesting to investigate whether such models can be seen as deformation of a free field theory in the spirit of Lechner's work, as well as to study the quantum group symmetry of the affine-Toda field theories.

\end{talk}

\end{document}